\documentclass[aps,prc,dlspace,twocolumn]{revtex4}
\usepackage{epsfig}
\usepackage{graphics}
\usepackage{color}

\newcommand{\jpsi}{J / \psi}

\newcommand{\old}[1]{}

\newcommand{\be}{\begin{equation}}
\newcommand{\ee}{\end{equation}}
\newcommand{\ba}{\begin{eqnarray}}
\newcommand{\ea}{\end{eqnarray}}

\begin{document}
\title{Dissociation of $1$p quarkonium states in a hot QCD medium}
\author{Vineet Agotiya$^1$, Vinod Chandra$^2$,
B. K. Patra$^1$\email{binoyfph@iitr.ernet.in}\\
$^1$ {\small Department of Physics, Indian Institute of 
Technology Roorkee, Roorkee-247 667, India}\\
$^2${\small Department of Physics,
Indian Institute of Technology Kanpur, Kanpur-208 016, India}}

\begin{abstract}
We extend the analysis of a very recent work 
(Phys. Rev. {\bf C 80}, 025210 (2009))
to study the dissociation phenomenon of $1$p states of the 
charmonium and bottomonium
spectra ($\chi_c$ and $\chi_b$) in a hot QCD medium. This study 
employed a medium modified heavy quark potential which is obtained by 
incorporating both perturbative and non-perturbative medium effects
encoded in the dielectric function to the full Cornell potential.
The medium modified potential has a quite different form
(a long range Coulomb tail in addition to the usual Yukawa term)
compared to the usual picture of Debye screening.
We further study the flavor dependence of their binding energies 
and dissociation temperatures by employing the perturbative, 
non-perturbative, and 
the lattice parametrized form of the Debye masses. 
These results are consistent with the predictions of the current 
theoretical works.
\end{abstract}
\maketitle

\section{Introduction}
One of the amazing discoveries of experimental measurements 
at RHIC is the surprising amount of 
both radial~\cite{expt} and elliptic flow~\cite{adler} exhibited
by the outgoing hadrons. 
Theoretical calculations cannot generate sufficient flow to 
explain the observations unless partonic cross sections are 
artificially enhanced by more than an order of magnitude 
over perturbative QCD predictions\cite{molnar1}.
Thus the matter created in these collisions
is strongly interacting, unlike the type of
weakly interacting quark-gluon plasma expected to occur at very 
high temperatures on the basis of asymptotic freedom~\cite{collin}.
The behavior of the heavy quarkonium states in hot strongly interacting matter
was proposed as test of its confinement status,
since a sufficiently hot deconfined medium will dissolve
any binding between the quark-antiquark pair \cite{matsui}.
Another possibility of dissociation of certain quarkonium states
(sub-threshold states at $T=0$) is the decay into open charm (beauty)
mesons due to in-medium modification of quarkonia and heavy-light
meson masses~\cite{Dig01}.

Many attempts have been made to understand the dissociation phenomenon  
of $Q\bar{Q}$ states in the deconfined medium, using either 
lattice calculations of quarkonium spectral 
functions \cite{Asa03,Dat04,Ume05,Jak07,gert} or non-relativistic
calculations based upon some effective  potential 
\cite{Shu04,Alb05,Won07,Cab06,Alb07,mocsyprd}.
These two approaches show poor matching between their predictions 
because of the uncertainties 
coming from a variety of sources. 
None of the approaches give a complete framework to study the 
properties of quarkonia states at finite temperature. 
However, some degree of qualitative agreement had been achieved 
for the S-wave correlators. The finding was somehow ambiguous for the
P-wave correlators and the temperature dependence of the potential model 
was even qualitatively different from the lattice one.
Refinement in the computations of the spectral functions have recently 
been done by incorporating the zero modes both in the S- and P-channels
\cite{Ume07,Alb08}.
It was shown that, these contributions cure most of the previously observed 
discrepancies with lattice calculations.
This supports the use of potential models at finite 
temperature as an important tool to complement lattice studies. 

The production of  $J/\psi$ and $\Upsilon$ mesons
in hadronic reactions occurs in part
through production of higher excited $c \bar c$ (or $b \bar b$) states
and their decay into quarkonia ground state. Since the lifetime of
different sub-threshold quarkonium states is much larger than the
typical life-time of the medium which may be produced in nucleus-nucleus
collisions; their decay occurs almost completely outside the produced
medium. This means that the produced medium can be probed not only by
the ground state quarkonium but also by different excited quarkonium states.
Since, different quarkonium states have different sizes (binding energies),
one expects that higher excited states will dissolve at smaller
temperature as compared to the smaller and more tightly bound ground states.
These facts may lead to a sequential suppression pattern in $J/\psi$ and
$\Upsilon$ yield in nucleus-nucleus collision as the function of the
energy density. So, if one wants to interpret the $J/\psi$ suppression pattern 
observed in nuclear collisions at CERN SPS and RHIC, as a signature 
of the formation of the QGP, one requires a right understanding of the 
dissociation of $\chi_c$
and $\chi_b$ in the QGP medium.
This is due to the fact that a significant fraction ($\sim 30 \%$)
of the $J/\psi$ yield observed in the collisions is produced by 
$\chi_c$ decays~\cite{Sat07,dpal,plb}. 
The $J/\psi$ yield could show a 
significant suppression even if the energy density of the system is 
not enough to melt directly produced $J/\psi$ but it
is sufficient to melt the higher resonance states
because they are loosely bound compared to the ground state $J/\psi$.
This motivates the special attention to the excited states $\chi_c$
and $\chi_b$.

In the studies of the bulk properties
of the QCD plasma phase~\cite{Kac1,Kac2,Bein,kars}, deviations 
from perturbative calculations
were found at temperatures much larger than the deconfinement
temperature. This calls for quantitative non-perturbative calculations.
The phase transition in full QCD appears as a crossover
rather than a `true' phase transition with related singularities in
thermodynamic observables (in the high-temperature and low density regime)
\cite{phaseT}. Therefore, it is not reasonable to assume that the string-tension
vanishes abruptly at or above $T_c$ and one should 
study its effect on the behavior 
of quarkonia even above the deconfinement temperature.
This issue, usually overlooked in the
literature, was certainly worth to investigate.
This is exactly what we have done in our recent work~\cite{akhilesh,v1} where 
we have obtained the medium-modified form of the heavy quark potential
by correcting the full Cornell potential (linear plus Coulomb), 
not only its Coulomb part alone as usually done in the literature, with
a dielectric function encoding the effects of the deconfined medium.
We found that this approach led to a long-range Coulomb potential 
with an (reduced) effective charge\cite{v1}  
in addition to the usual Debye-screened form employed in most
of the literature. With this effective potential, we investigated the
effects of perturbative and non-perturbative contributions
to the Debye mass on the dissociation of quarkonium states. 
We subsequently used this study to determine
the binding energies and the dissociation temperatures of the ground and 
the first excited states of charmonium and bottomonium spectra.

However, our starting potential (Cornell) at $T=0$ has 
no terms to account
the spin-dependence forces in QCD~\cite{Eichten}, so the medium-modified 
potential~\cite{v1} also has no spin-dependent terms.
As a consequence, Schr\"odinger equation
with the above medium-modified potential gives 
the same energy eigenvalues for 
the first excited states $\psi^\prime$ and $\chi_c$ 
making them degenerate. This is certainly not desirable since  
their masses are not the same (in fact, mass of $\psi^\prime$ is 
slightly higher than $\chi_c$ whereas the latter state is more tightly bound
than the former).
Therefore the determination of the binding energies of 1p
states, {\it viz.}, $\chi_c$, $\chi_b$ 
and their dissociation temperatures like the ground  and first excited 
states is 
not directly possible as had been done in our earlier work
by employing the medium modified potential~\cite{v1}.
The principal quantum number ($n$) of $\psi^\prime$ and $\chi_c$ are 
same
but their spin quantum number and as well as their total angular momentum
are not the same.  So, their quantum states should be denoted by
all four quantum numbers ($nlsj$) and the difference in their 
binding energies
(or in their total masses) should be originated from a spin-dependent 
correction terms.

We have done this job in a two fold way.  First, 
we have determined the binding energy for $\psi^\prime$
by employing the medium-modified potential~\cite{v1} into
the Schr\"odinger equation. Then
we obtain the binding energy for $\chi_c$ by adding  
the correction terms to the binding energy of
$\psi^\prime$. In our  analysis,  correction terms will be obtained 
by adopting a variational treatment of the 
relativistic two-fermion bound-states in quantum 
electrodynamics (QED)~\cite{Dare1,Dare3} taking into account the spin-dependent
terms for the corresponding quantum numbers 
of $\psi^\prime$ and $\chi_c$ states.
In this endeavor, coupled integral equations for a relativistic two-fermion
system are derived variationally within the Hamiltonian formalism of
QED using an improved ansatz that is sensitive
to all terms in the Hamiltonian~\cite{Dare1}. 

The paper is organized as follows. In Sec.II, we review the work on
the medium modified Cornell potential and dissociation of $1 s$ and $2 s$
states of charmonium and bottomonium spectra. In Sec.III, we discuss how 
to determine the binding energies of $\chi_c$ and $\chi_b$. In Sec.IV, 
we study the melting of $\chi_c$ and $\chi_b$ in the QGP medium and
determine their dissociation temperatures. 
Finally, we conclude in Sec.V. 

\section{In-medium modifications to heavy-quark potential}
The interaction potential between a heavy quark and antiquark gets
modified in the presence of a medium and it
plays a vital role in understanding the fate of
quark-antiquark bound states in the QGP medium.
This issue has well been studied and several excellent
reviews exist~\cite{Bram,kluberg} which dwell both on
the phenomenology as well as on the lattice QCD. In these studies, they
assumed the melting of the string motivated by the fact
that there is a phase transition from a hadronic matter to a QGP phase.
As a consequence they modified the Coulomb part of the potential only
so they used a much simpler form (screened Coulomb) of the medium 
modified potential in the deconfined phase.
But recent lattice results indicates that there is no genuine
phase transition at vanishing baryon density, it is rather
a cross-over, so there is no reason to assume the melting of string
at the deconfinement temperature. We have addressed this issue in our
recent work~\cite{v1} where we  
developed an effective potential once one corrects
the full Cornell potential with a dielectric function embodying medium
effects.  We recall the basic details
which are relevant for the present demonstration.

Usually, in finite-temperature QFT, medium modification enters in the 
Fourier transform of heavy quark  potential as 
\begin{equation}
\label{eq3}
%\tilde {\bf V}(k)=\frac{{\matcal V}(k)}{\epsilon(k)}
\tilde{V}(k)=\frac{V(k)}{\epsilon(k)} \quad ,
\end{equation}
where $\epsilon(k)$ is the dielectric permittivity given in terms
of the static limit of the longitudinal part of gluon
self-energy\cite{schneider}:
\begin{eqnarray}
\label{eqn4}
\epsilon(k)=\left(1+\frac{ \Pi_L (0,k,T)}{k^2}\right)\equiv
\left( 1+ \frac{m_D^2}{k^2} \right).
\end{eqnarray}
The quantity  $V(k)$ in Eq.(\ref{eq3}) is the Fourier transform (FT) of
the Cornell potential. The evaluation of
the FT of the Cornell potential is not so straightforward
and can be done by assuming $r$- as distribution
($r \rightarrow$ $r \exp(-\gamma r))$. After the evaluation of
FT we let $\gamma$ tends to zero. 
Now the FT of the full Cornell potential can be written as
\begin{equation}
\label{eqn5}
{\bf V}(k)=-\sqrt{(2/\pi)} \frac{\alpha}{k^2}-\frac{4\sigma}{\sqrt{2}\pi k^4}.
\end{equation}
Substituting Eqs.(\ref{eqn4}) and (\ref{eqn5}) into (\ref{eq3})
and then evaluating its inverse FT
one obtains the r-dependence of the medium modified
potential~\cite{akhilesh}:
\begin{eqnarray}
\label{eq4}
{\bf V}(r,T)&=&\left( \frac{2\sigma}{m^2_D}-\alpha
\right)\frac{\exp{(-m_Dr)}}{r}\nonumber\\
&-&\frac{2\sigma}{m^2_Dr}+\frac{2\sigma}{m_D}-\alpha m_D
\end{eqnarray}
This potential has a long range Coulombic tail in addition to the
standard Yukawa term. 
The constant terms 
are introduced  to yield the correct limit of $V(r,T)$ as $T\rightarrow 0$ 
(it should reduce to the Cornell form).
Such terms could arise
naturally from the basic computations of real time static potential 
in hot QCD\cite{const1} and from the real and imaginary time
correlators in a thermal QCD medium\cite{const2}.

It is worth to note that the potential
in a hot QCD medium is not the same as the lattice parametrized heavy
quark free-energy in the deconfined phase which is basically a screened
Coulomb~\cite{hsatz,shuryak} because one-dimensional Fourier
transform of the Cornell potential in the medium yields the similar form
as used in the lattice QCD to study the quarkonium properties
which assumes the one-dimensional color flux tube structure~\cite{dixit}.
However, at finite temperature that may not be
the case since the flux tube structure may expand in more
dimensions\cite{hsatz}.
Therefore, it is better to consider the three-dimensional form of the medium
modified Cornell potential which have been done exactly in the present work.
We have compared our in-medium potential with the color-singlet 
free-energy\cite{pot_1} extracted from the lattice data 
and found that it agrees with the lattice results except 
from the non-perturbative result of the Debye masses.

However, if we neglect the finite range terms {\em viz.} Yukawa term 
in the limit $r>>1/m_D$
and for large values of temperatures the product
$\alpha m_D$ will be much greater than $2\sigma/m_D$,
then it leads to an analytically solvable Coulomb potential:
\begin{eqnarray}
\label{lrp}
{V(r,T)}\sim -\frac{2\sigma}{m^2_Dr}-\alpha m_D
\end{eqnarray}
We employed this medium-modified 
effective potential to study the 
binding energies and the dissociation temperatures for 
the ground and first excited states of $c \bar c$ and $ b \bar b$ 
spectroscopy. However, to see the effects of the finite-range terms, 
we solve the Schr\"odinger equation numerically with the full effective
potential(\ref{eq4}) and found that the dissociation temperatures was 
changed by  $\sim$ 10\%~\cite{v1}. 
Therefore the approximated form
(\ref{lrp}) have a dominant role in deciding the fate of these states in the
hot QCD medium. Let us now proceed to the determination of the binding energies and the 
dissociation temperatures for $\chi_c$ and $\chi_b$ states in Sec(s).III 
and IV, respectively.

\section{Binding energy of $\chi_c$ and $\chi_b$}
The in-medium potential (\ref{lrp}) resembles to the hydrogen 
atom problem. The solution of
the Schr\"{o}dinger equation gives the eigenvalues for the 
ground states and the first excited states 
in charmonium ($\jpsi$, $\psi^\prime$ etc.) and 
bottomonium ($\Upsilon$, $\Upsilon^\prime$ etc.) spectra :
\begin{eqnarray}
\label{bind1}
E_n=-\frac{E_I}{n^2} \quad; \quad E_I=\frac{m_Q\sigma^2}{m^4_D},
\end{eqnarray}
where $m_Q$ is the mass of the heavy quark and $E_I$ is the energy of the
$Q\bar{Q}$ state in its first Bohr orbit. 
The allowed energy states 
for $Q\bar{Q}$  are $E_n=-E_I, -\frac{E_I}{4}, 
\cdot \cdot \cdot$.
These energies are known as the ionization
potentials/binding energies for the $n$th bound state. 
They become temperature-dependent through
the Debye masses and decrease with the increase in temperature. 

Apart from the ground and the first excited states, 
there are other important states ($1$p) 
in the charmonium and bottomonium 
spectra {\it viz} $\chi_c$ and $\chi_b$
which contribute significantly in the suppression of ground state 
quarkonia ($J/\psi$ and $\Upsilon$)
in RHIC experiments through their decays into $J/\psi$'s and 
$\Upsilon$'s. Although both $\psi^\prime$ and $\chi_c$ are 
the first excited states of the charmonium spectra 
but they are not degenerate. In fact, $\psi^\prime$ is more massive
than $\chi_c$ but $\chi_c$ is more
tightly bound than $\psi^\prime$. So the
entire binding energy of $\chi_c$ will not come from the above
calculation, the additional contribution will come from the
spin-dependent quantum corrections.

Some authors have studied the relativistic two-particle
Coulomb problem, based on approximations to the Bether-Salpter equations.
Others have started with effective Lagrangians based on perturbative
expansions of the relativistic Lagrangians. However, we choose
the variational methods~\cite{hard,Dare1,Dare3} where
coupled integral equations for a relativistic two-fermion
system are derived variationally 
within the Hamiltonian formalism of quantum electrodynamics,
using an improved ansatz that is sensitive
to all terms in the Hamiltonian. 
The equations are solved approximately to determine the eigenvalues
and eigenfunctions, at arbitrary coupling, for various states of the
two-particle system.
In the variational treatment of the relativistic two-fermion
bound-state system in QED, the total energy
in a quantum state ($nlsj$) consists of 
Bohr like terms, relativistic correction in the
kinetic energy and most importantly the spin-dependent terms
which take into account the non-degeneracy between the sub-states.
The total energy up to fourth order in $\alpha$ 
is written for a hydrogen-like potential~\cite{Dare1} 
\be
\label{chi}
E_{nlsj}=2m_Q -\frac{1}{2}\mu \frac{\alpha^2}{n^2} +\Delta K_{nl}
+\Delta V_{nlsj}~, 
\ee
where
\be
\Delta K_{nl}= -\frac{\mu^4 \alpha^4}{m_Q^3} \left( \frac{2}{(2l+1)n^3}
- \frac{3}{4n^4} \right) 
\ee
is the $\alpha^4$ correction to the kinetic energy and the correction
to the spin-dependent potential energy is 
\be
\Delta V_{nlsj}= - b{{}_{nlsj}} \frac{ \alpha^4 \mu^3}{n^3} ~,
\ee
where the coefficients $b_{{}_{nlsj}}$ 
for the different quantum states ($nlsj$) are tabulated in 
Ref.\cite{Dare1}. We have taken their values for 
the Coulombic case only. The fine structure constant ($\alpha$) in
QED will be replaced by the effective charge 
($2\sigma/m_D^2$) in  our model. Using the appropriate 
values of the quantum numbers and the coefficients 
corresponding to $\psi^\prime$ and $\chi_c$ states in charmonium
spectra ($\Upsilon^\prime$ and $\chi_b$ in bottomonium spectra), we 
obtain the correction term which is to be added to the 
binding energy of $\psi^\prime$ is
\be
E^{\mbox{corr}}_{c,b}=\frac{m_{c,b}\alpha^4}{96}=\frac{m_{c,b}\sigma^4}{
6m_D^8}
\ee
So the binding energy of $\chi_c(\chi_b)$ is
\ba
\label{bind2}
E(\chi_{{}_{c}},\chi_{{}_{b}}) 
&=& E(\psi^\prime,\Upsilon^\prime) +E^{\mbox{corr}}_{\chi_c,\chi_b}
\nonumber\\
&=& \frac{m_{{}_{c,b}}\sigma^2}{4 m^4_D} \left( 1 + \frac{2}{3}
 \frac{\sigma^2}{m^4_D} \right)~,
\ea
where $m_D$ is the 
Debye mass for which we choose
a gauge invariant, non-perturbative form~\cite{kaj1}.
Recently Kajantie {\it et. al}~\cite{kaj1} obtained it by
computing the non-perturbative contributions of ${\cal O}(g^2T)$ and
${\cal O}(g^3T)$ from a 3-D effective field theory as
\begin{eqnarray}
\label{eq}
m^{\mbox{NP}}_D &=& m^{\mbox{LO}}_D 
+{N_c g^2T\over4\pi}\ln{m^{\mbox{LO}}_D\over  g^2T} \nonumber\\
&+&  c_{{}_{N_c}} g^2T + d_{{}_{{N_c},N_f}} g^3 T + {\cal O}(g^4T) \quad,
\end{eqnarray}
where the leading order (LO) perturbative result, 
$m^{\mbox{LO}}_D = g(T) T \sqrt{\frac{N_c}{3}+ \frac{N_f}{6}}$,
has been known for a long time~\cite{shur}. The logarithmic part of the
${\cal O}(g^2)$ correction can be extracted perturbatively~\cite{rebh}, but
$c_{N_c}$ and the higher order corrections are non-perturbative.
We wish to explore the effects of the different terms in the
Debye mass on the binding energy of $\chi_c$ and $\chi_b$.
We have used the two-loop expression for the QCD 
coupling constant at finite temperature from Ref.\cite{shro} and 
the renormalization scale from Ref.\cite{shaung}.

The effects of each terms in the Debye mass (\ref{eq}) cannot
always be explored separately due to the following reason:
In the weak coupling
regime, the soft scale ($\simeq gT$)
at the leading-order related to the screening of electrostatic fields
is well separated from the ultra-soft scale ($\simeq g^2T$)
related to the screening of magnetostatic fields. In such regime,
it appears meaningful to see the contribution of each terms
in the Debye mass separately. But when the
coupling becomes large enough (which is indeed the case), the
two scales are no longer
well separated. So while looking for the next-to-leading corrections
to the leading-order result from the ultra-soft scale, it is not a wise
idea to stop at the logarithmic term, since it becomes crucial the 
number multiplying the
factor $1/g$ to establish the correction to the LO result.
In fact the Debye mass in the NLO term
is always smaller than the LO term 
because of the negative (logarithmic) contribution ($\log (1/g)$)
to the leading-order term, while the full correction (all $g^2T$ terms)
to the Debye mass results positive. So, we consider only three forms of the
Debye masses, {\it viz.} leading-order result ($m_D^{\rm{LO}}$), 
non-perturbative
form ($m_D^{\rm{NP}}$), and lattice parametrized
form ($m_D^L=1.4 m_D^{\rm{LO}}$) to study the dissociation 
phenomena.

\begin{figure*}
\vspace{10mm}
\includegraphics[scale=.30]{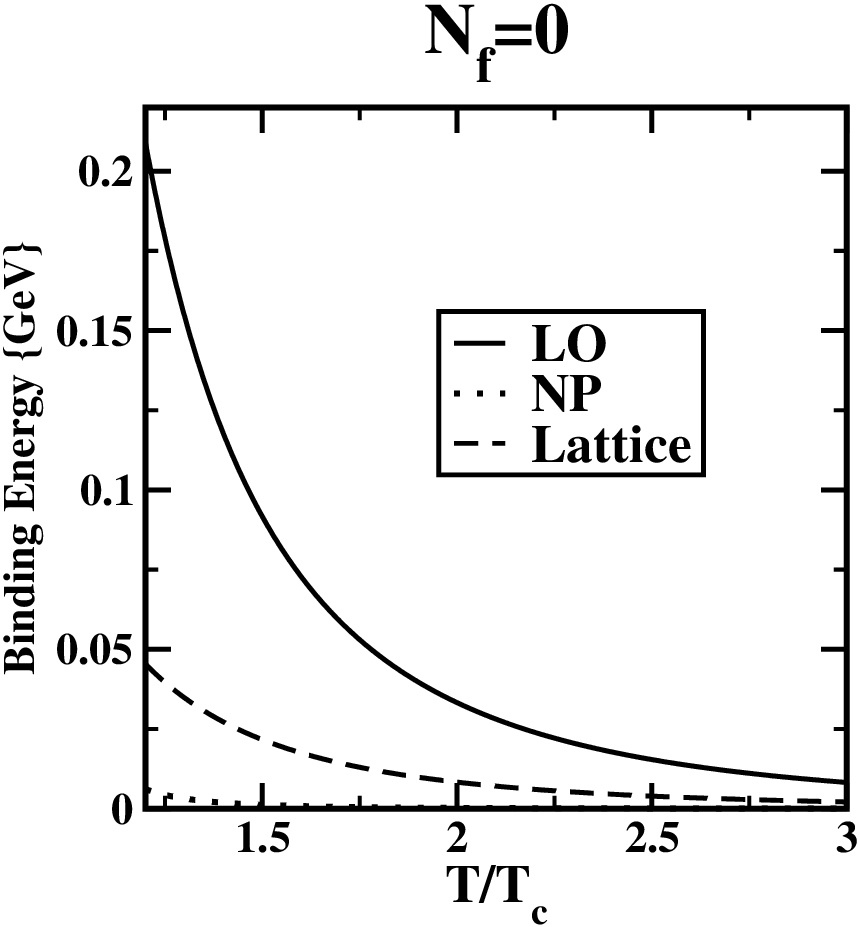} % Here is how to import EPS art
\hspace{2mm}
\includegraphics[scale=.30]{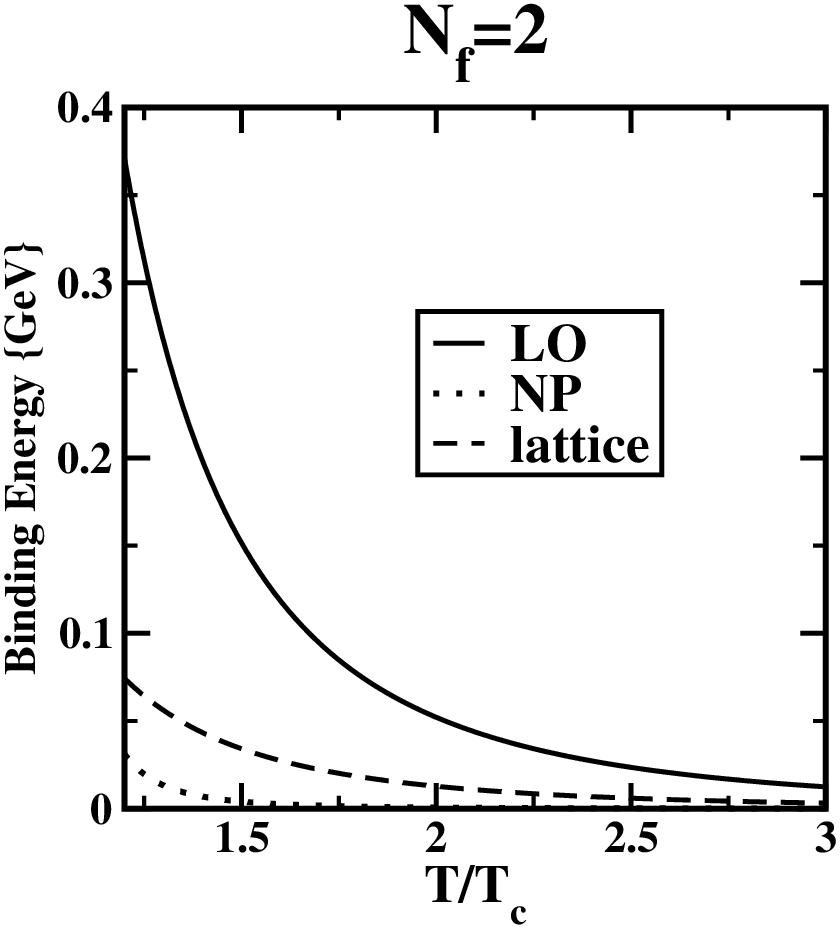}
\hspace{2mm}
\includegraphics[scale=.30]{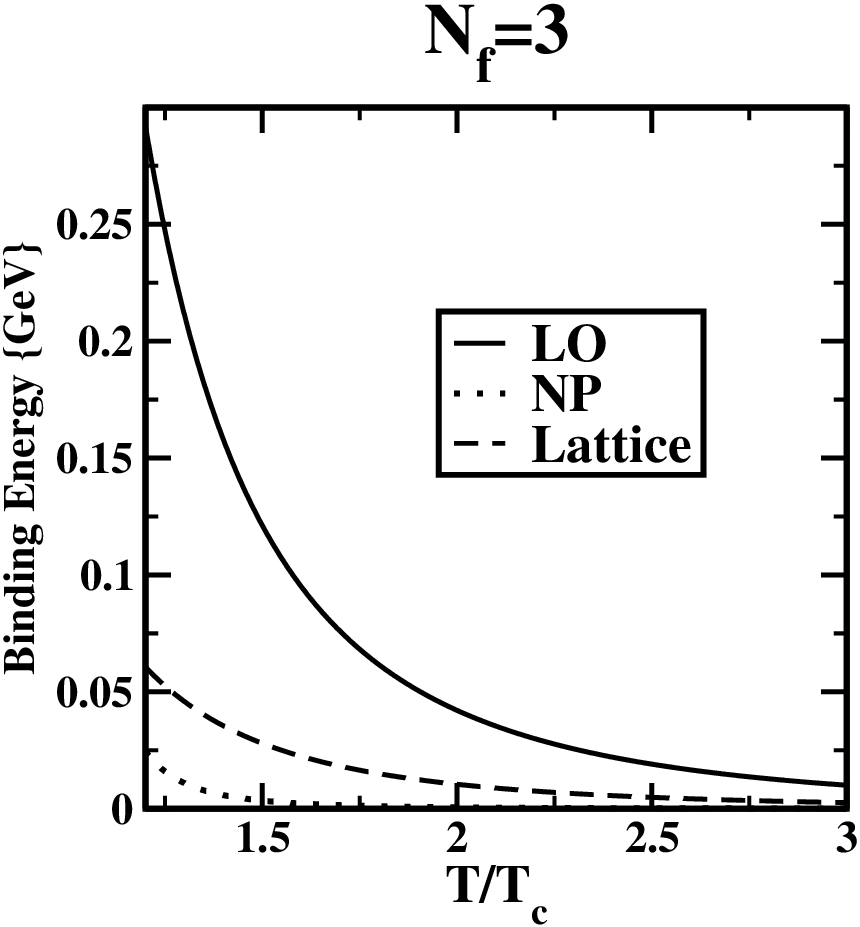}
\caption{Dependence of $\chi_c$ binding energy (in $GeV$) on 
temperature $T/T_c$.}
\vspace{10mm}
\end{figure*}

Thus we have finally computed the binding energies 
for $\chi_c$ and $\chi_b$ and plotted them in Figs. 1 and 2,
respectively where different curves denote the choice of 
the Debye masses used to calculate the binding energy from 
Eq.(\ref{bind2}). We consider three cases for 
our analysis: pure gluonic, 2-flavor and 3-flavor QCD.
There is a common observation in all figures that 
the binding energies show strong decrease with increase in temperature.
In particular, binding energies obtained from 
$m^{LO}_D$ and $m^L_D$ give realistic variation with the 
temperature. The temperature dependence of the binding energies show 
a quantitative agreement with the results based on the  
spectral function technique calculated in a potential model for
the non-relativistic Green's function \cite{mocsyprl}.
On the other hand, when we employ non-perturbative form of the
Debye mass ($m^{NP}$) the binding energies
become unrealistically small compared to 
its zero temperature value and also compared to the binding 
energies employing  $m^{LO}_D$ and $m^{L}_D$.
This anomaly can be understood by the fact that the value of $m^{NP}_D$
is significantly larger than both $m^{LO}_D$ and $m^{L}_D$
so that the binding energies become substantially smaller.
This observation indicates that the present 
form of the non-perturbative corrections to the Debye mass 
may not be the complete one, the situation may change 
once the $O(g^4T)$ non-perturbative contributions to Debye mass are 
incorporated and then evaluate the binding energy. 
Thus, the study of temperature dependence of binding energy is poised
to provide a wealth of information about the nature of dissociation
of quarkonium states in a thermal medium which will be 
reflected in their dissociation temperatures 
discussed in the next section.

In addition, we take advantage of all
the available lattice data, obtained not only in
quenched QCD ($N_f=0$), but also including two and, more recently, three light
flavors. We are then in a position to study also the flavor dependence of the
dissociation process, a perspective not yet achieved by the parallel studies of
the spectral functions, which are only available in
quenched QCD.

\section{Dissociation temperatures}
It has been customary to consider a state dissociated when its binding
energy becomes zero. In principle, a state is dissociated when no peak
structure is seen, but the widths shown in spectral functions from
current potential model calculations are not physical.
Broadening of states as the temperature increases is not included in any
of these models. In~\cite{mocsyprl} authors have argued that no need
to  reach zero binding energy ($E_{\rm{bin}}=0$) to dissociate, but
when $E_{\rm{bin}} \le T$ a state is weakly bound and thermal
fluctuations can destroy it. However, others have set a more conservative 
condition for dissociation~\cite{dima}: $2 E_{bin}(T)< \Gamma(T)$, where 
$\Gamma(T)$ is the thermal width of state. However, we now calculate the upper 
bound of the dissociation temperature by the condition for 
dissociation: $|E_{\chi_{c,b}}| \approx T$,
\begin{figure*}
\vspace{10mm}
\includegraphics[scale=.30]{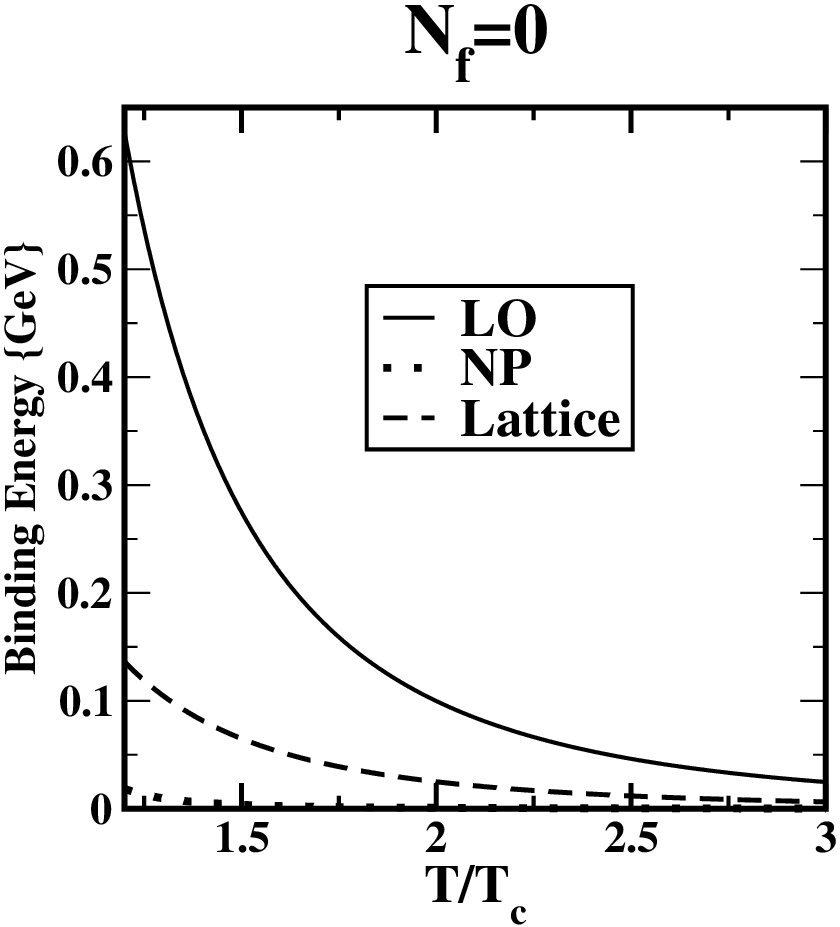} % Here is how to import EPS art
\hspace{2mm}
\includegraphics[scale=.30]{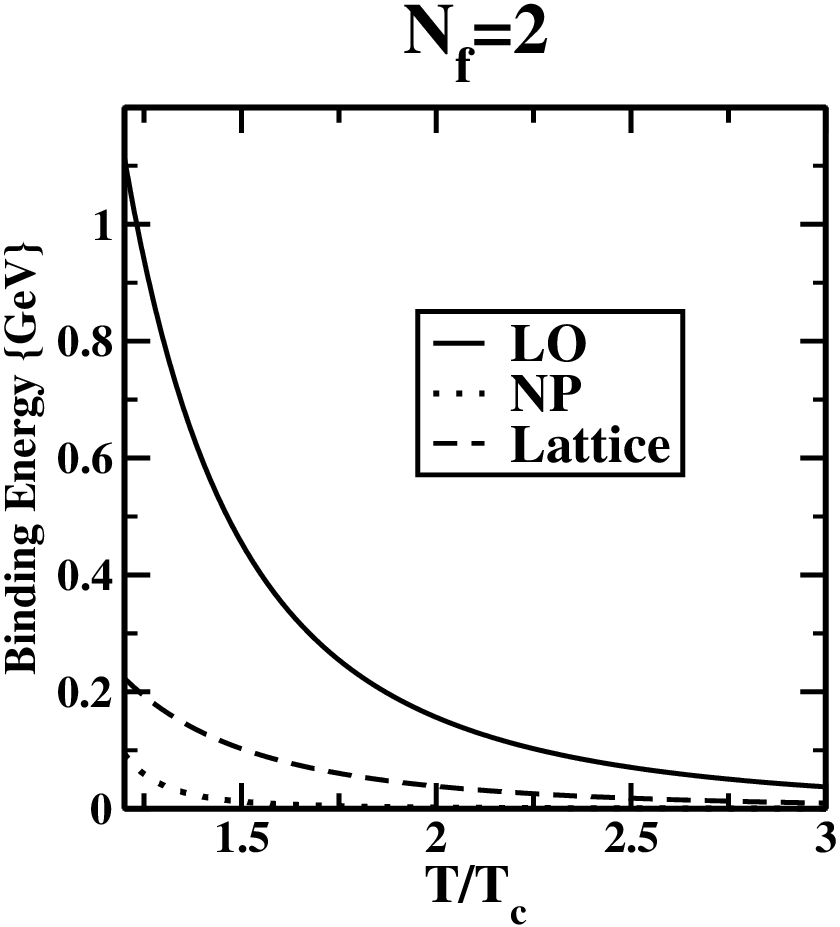}
\hspace{2mm}
\includegraphics[scale=.30]{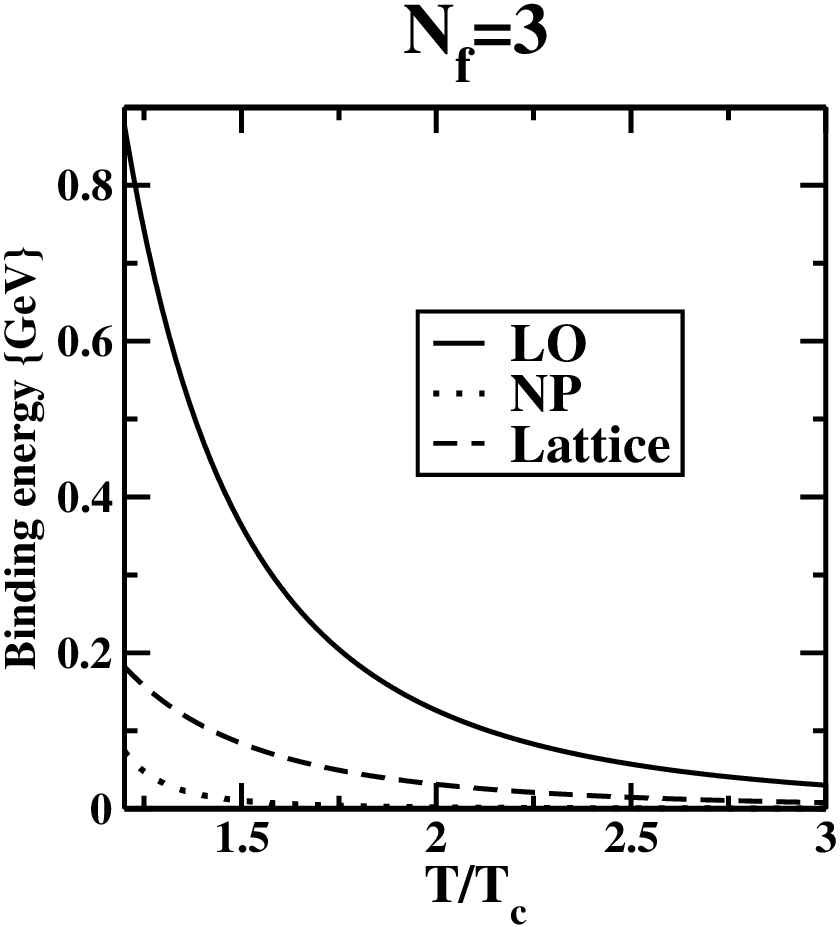}
\caption{Dependence of $\chi_b$ binding energy  (in $GeV$) on $T/T_c$.}
\vspace{15mm}
\end{figure*}

\begin{equation}
\label{tdiss}
E(\chi_{{}_{c,b}}) = \frac{m_{{}_{c,b}}\sigma^2}{4 m^4_D} 
\left( 1 + \frac{2}{3}
 \frac{\sigma^2}{m^4_D} \right)= T_D ~,
\end{equation}
where the string tension ($\sigma$) is 0.184 ${\rm{GeV}^2}$ 
and critical temperatures ($T_c$) are taken as
$270 MeV$, $203 MeV$ and $197 MeV$ for pure gluonic, 
2-flavor and 3-flavor QCD medium, respectively\cite{zantow}.
However, the choice of the mean thermal energy ($T$) is not rigid 
because even at low temperatures
$T < T_D$ (say) the Bose/Fermi distributions of partons will 
have a high energy tail with partons of mechanical energy 
$> |E_{\chi_{c,b}}|$. 
\begin{table}
\caption{\label{table1} Upper(lower) bound on the dissociation 
temperature($T_D$) for $\chi_c$ and $\chi_b$ (in unit of $T_c$)
using the leading-order term in Debye mass, $m^{LO}_D$.}
\centering
\begin{tabular}{|l|l|l|l|l|}
\hline
State &Pure QCD & $N_f=2$&$N_f=3$\\
\hline
$\chi_c$&1.10 (0.89) &1.31 (1.06) &1.26 (1.02) \\
\hline\hline
$\chi_b$&1.38 (1.10)   &1.64 (1.31)  &1.57 (1.26) \\
\hline
\end{tabular}
\end{table}
The dissociation temperatures for 
$\chi_c$ and $\chi_b$ 
are listed in Table I with the Debye mass in the leading-order.
It is found that $\chi_c$'s are dissociated at $1.1 T_c$, $1.31T_c$,
and $1.26T_c$ for the pure, 2-flavor,
and 3-flavor QCD, respectively whereas
$\chi_b$'s are dissociated comparatively much higher
temperature which seems justifiable.
This is perhaps the first observation in the literature 
on the flavor (system) dependence
of the dissociation temperature. This dependence is essential
while calculating the screening energy density (energy density
at the dissociation temperature) in various descriptions
of QGP ($N_f=0,2,3$) for the study of $J/\psi$ survival in 
an expanding QGP. On the other hand, employing 
lattice parametrized form, $m^{L}_D$ we obtain the 
values (Table II) much smaller than the 
leading-order results where $\chi_c$ is dissociated 
below $T_c$ and $\chi_b$ is dissociated just above $T_c$. 
At last, when we use non-perturbative form of the Debye mass,
the dissociation temperatures come out to be unrealistically small.
Summarizing the results, we conclude that as we move from 
perturbative to non-perturbative domain, the binding
energies are becoming smaller and smaller. As a result the
dissociation temperatures obtained are also becoming
smaller. This is due to
the hierarchy in the Debye masses: $m^{\mbox{LO}}_D <
m^{\mbox{L}}_D \ll m^{\mbox{NP}}_D$. In fact, $m^{L}_D$ is
1.4 times greater than $m^{\mbox{LO}}_D$ while $m^{\mbox{NP}}_D$ 
is much greater than both $m^{LO}_D$ and $m^{L}_D$.
\begin{table}
\caption{\label{table2} Same as Table I but with
the lattice parametrized form of the Debye 
mass $m^{L}_D$.}
\centering
\begin{tabular}{|l|l|l|l|l|}
\hline
 State &Pure QCD & $N_f=2$&$N_f=3$\\
\hline\hline
$\chi_c$&0.77 (0.59) &0.93 (0.73) &0.90 (0.71) \\
\hline
$\chi_b$&0.98 (0.77) &1.17 (0.93) &1.14 (0.90) \\
\hline
\end{tabular}
\end{table}

\begin{table}
\caption{\label{table3} Upper bound of the dissociation temperatures 
with $T_c=192$ MeV and the lattice parametrized form of the 
Debye mass~\cite{mocsyprl}.}
\centering
\begin{tabular}{|c|c|c|}
\hline
 State & $\chi_c$ & $\chi_b$\\
\hline
$T_D$ & $\le 0.9T_c$ & 1.2$T_c$ \\
\hline
\end{tabular}
\end{table}
However, if we treat the partons in high temperature to be relativistic,
we could replace the mean thermal energy by $3T$ (instead of $T$) 
to obtain the lower bound for the dissociation temperatures. 
It is found that all entries in Table I and II have been decreased by 
30\% approximately giving the lower bound of the dissociation temperatures 
(inside the first bracket). 
To compare our results quantitatively 
with the recent results~\cite{mocsyprl}
based on the spectral function technique calculated in a potential model
with a similar description of the system (for 3-flavor QCD with $T_c$=192 MeV), 
we tabulated the upper limit on the dissociation temperatures with the 
same form of Debye mass used in Ref.\cite{mocsyprl} 
in Table III giving a good agreement with their results. 

Finally, it is learnt that inclusion of non-perturbative corrections 
to the Debye mass ($m^{NP}_D$)
leads to unusually smaller value of the dissociation temperatures 
for both $\chi_{c}$ and $\chi_{b}$. This does not immediately imply that
the non-perturbative effects should be ignored. It is rather interesting 
to investigate the disagreement between the non perturbative result 
obtained with a dimensional-reduction strategy and the Debye mass
arising from the Polyakov-loop correlators.
Only future investigation may throw more light on this 
issue. 

\section{Conclusions} 
In conclusion, we have studied the dissociation of  $1 p$ states 
in the charmonium and bottomonium spectra($\chi_c$ and $\chi_b$)
in the hot QCD medium. We have employed the medium 
modified form of the heavy quark-potential in which the
medium modification causes the dynamical screening
of color charge which, in turn, leads to the temperature dependent 
binding energy of $\psi^\prime$ and $\Upsilon^\prime$. We have then
studied the temperature dependence of the binding energy of the 
$\chi_{{}_{c}}$ and $\chi_{{}_{b}}$ 
states in the pure gauge  and realistic QCD medium 
by incorporating the fourth-order corrections (in the screened 
charge $\alpha_{\mbox{eff}}= 2 \sigma/m^2_D$)
coming from the spin dependent terms to
the binding energies of $\psi^\prime$ and $\Upsilon^\prime$ states,
respectively. For this purpose, we have 
adopted a formulation~\cite{Dare1},
in which a variational treatment of the relativistic
two-fermion bound-state system in QED~\cite{Dare1} has been
developed to compute the spin-dependent corrections.

Next we have determined the dissociation temperatures employing the 
Debye mass in leading-order and  
the lattice parametrized form. Our estimates are 
consistent with the finding of recent theoretical works based on 
potential models~\cite{mocsyprl}. We have further shown that 
inclusion of non-perturbative contributions to the Debye mass 
lower the dissociation temperatures substantially which looks 
unfeasible to compare to the spectral
analysis of lattice temporal correlator of  mesonic current.
This leaves an open problem of the agreement between
these two kind of approaches.
This could be partially due to the arbitrariness in the 
criteria/definition of the
dissociation temperature.
To examine this point we have estimated both the upper and lower 
bound on the dissociation temperatures by fixing the
mean thermal energy $T$ and $3T$, respectively.
Thus, this study provides us a handle to decipher 
the extent upto which non-perturbative effects should 
be incorporated into the Debye mass.

In brief, we obtained the analytic forms for the 
binding energies and the dissociation temperatures of 
$\chi_c$ and $\chi_b$. This enable us to 
investigate their flavor dependence and temperature dependence.
We have estimated the upper bound on the dissociation temperatures
of $\chi_c$ and $\chi_b$. We found that these estimates obtained 
by employing the lattice parametrized Debye mass 
show good agreement with the prediction in \cite{mocsyprl}. 
On the other hand, these values are significantly smaller than the 
predictions of lattice studies~\cite{Dat04,Alb07,gert,Sat07}.

\end{document}